\documentclass[journal]{IEEEtran}
\usepackage{newunicodechar}
\newunicodechar{∗}{*}
\usepackage{tabularx}
\usepackage{cite}
\usepackage{amsmath,amssymb,amsfonts}
\usepackage{graphicx}
\usepackage{setspace}
\usepackage{textcomp}
\usepackage{xcolor}
\usepackage{float}
\usepackage{fancyhdr}
\usepackage{multicol}
\usepackage{setspace}
\usepackage{algorithm}
\usepackage[noend]{algorithmic}
\usepackage{hyperref}
\usepackage{enumitem}
\usepackage{todonotes}
\usepackage{comment}
\usepackage[normalem]{ulem}
 \usepackage{multirow} 
\usepackage{amssymb}
\usepackage{pifont}
\newcommand{\cmark}{\ding{51}}  
\newcommand{\xmark}{\ding{55}}  

\newenvironment{thisnote}{\par\color{blue}}{\par}

\def\BibTeX{{\rm B\kern-.05em{\sc i\kern-.025em b}\kern-.08em
    T\kern-.1667em\lower.7ex\hbox{E}\kern-.125emX}}
\begin{document}

\title{DSPE: Profit Maximization in Edge-Cloud Storage System using Dynamic Space Partitioning with Erasure Code}


\author{\IEEEauthorblockN{Shubhradeep Roy, Suvarthi Sarkar, Vivek Verma, Aryabartta Sahu }
\IEEEauthorblockA{Indian Institute of Technology Guwahati, Emails: \textit{\{shubhradeep.roy, s.sarkar, v.verma, asahu\}@iitg.ac.in }} 
}

\maketitle

\pagestyle{fancy}
\pagenumbering{arabic}

\begin{abstract}
Edge Storage Systems have emerged as a critical enabler of low-latency data access in modern cloud networks by bringing storage and computation closer to end-users. However, the limited storage capacity of edge servers poses significant challenges in handling high-volume and latency-sensitive data access requests, particularly under dynamic workloads.
In this work, we propose a profit-driven framework that integrates three key mechanisms: \textit{collaborative caching}, \textit{erasure coding}, and \textit{elastic storage partitioning}. Unlike traditional replication, erasure coding enables space-efficient redundancy, allowing data to be reconstructed from any subset of $K$ out of $K+M$ coded blocks. We dynamically partition each edge server’s storage into \textit{private} and \textit{public} regions. The private region is further subdivided among access points (APs) based on their incoming request rates, enabling adaptive control over data locality and ownership.

We design a data placement and replacement policy that determines how and where to store or evict coded data blocks to maximize data access within deadlines. While the private region serves requests from local APs, the public region handles cooperative storage requests from neighboring servers.
Our proposed Dynamic Space Partitioning and Elastic (DSPE) caching strategy is evaluated on both synthetic and real-world traces from Netflix and Spotify. Experimental results show that our method improves overall system profitability by approximately \textit{5--8\%} compared to state-of-the-art approaches under varied workload conditions.
\end{abstract}

 \section{Introduction}

 The exponential growth of data and the surge in Internet-connected devices have reshaped the modern computing environment. To address the rising need for fast and efficient services, enterprises are increasingly adopting cloud-based storage and computing solutions. This shift is largely driven by the shortcomings of traditional centralized data centers, which struggle to meet the scalability and latency demands of modern applications \cite{gartner_edge}. The move toward decentralization has led to greater internet congestion and a higher risk of service disruptions.
 
Data-intensive applications like video streaming, online retail, and real-time analytics often operate under soft deadline constraints, typically requiring response times under 300 milliseconds\cite{Liu20}. Studies show that even slight latency increases can lead to notable performance and revenue losses—for instance, a 100 ms delay on Amazon can reduce sales by up to 1\%\cite{Liu20}. To address this, cloud and internet service providers have heavily invested in infrastructure to reduce latency and maintain service quality. However, efficiently handling large-scale, real-time data while meeting tight deadlines remains a significant challenge.

 Edge computing has emerged as a promising alternative to centralized architectures, addressing key challenges by utilizing a distributed network of edge servers (ESs) equipped with storage and computational capabilities. Positioned close to end-users, these ESs reduce latency, optimize bandwidth usage, and enhance service quality. For instance, Ren \textit{et al.} \cite{a3} studied both dense (70 ESs) and sparse (15 ESs) deployments within a $10\times10$ grid, where servers were interconnected via high-speed links to form a cooperative Edge Storage System (ESS). 
 Due to the limited storage capacity of edge servers (ESs) and their inability to handle large volumes of user requests independently, traditional caching strategies relying solely on replacement policies may fall short. To overcome these constraints, a \textit{collaborative caching} approach is employed, where ESs coordinate and exchange data blocks with neighboring servers. This cooperation allows for a larger cumulative cache across the network, enhancing the system’s ability to serve more user requests efficiently.

\textit{Erasure coding} has proven to be an efficient method for reducing storage overhead and costs in cloud-based storage systems\cite{jin2022cost}. This technique involves dividing the original data 
$X$ into $K$ data blocks and 
$M$ parity blocks, which are then distributed across multiple storage nodes, such as edge servers (ESs), accessible to users. Data reconstruction requires retrieving any 
$K$ distinct blocks from the network. However, erasure coding introduces three major constraints: the proximity constraint\cite{a4}, concerning the spatial distance between users and ESs; the encoding constraint\cite{xia21}, related to the computational cost of encoding; and the transmission constraint\cite{a5}, involving bandwidth and latency of data transfer. Jin \textit{et al.} \cite{jin2022cost} effectively addressed these limitations by proposing a comprehensive framework that integrates all three constraints.

Building upon prior research on data placement in ESs utilizing erasure coding, we introduce a novel coalition-based \textit{Dynamic Space Partitining} policy that partitions the storage space of each ES into two distinct segments: \textit{private} and \textit{shared (public)}. The private segment is reserved for storing and replacing content requested directly by users through associated access points (APs). Portion of the private space allocated to each AP is dynamic and changes according to the change in request arrival rate in each AP. Whereas the shared segment is allocated for storing and replacing content requested by users through neighboring ESs. This elastic caching strategy is designed to enhance system performance by effectively balancing local and global caching needs.
The rationale behind this partitioned design is:

(a) In a fully shared storage configuration, an ES experiencing a disproportionately high volume of user requests relative to its peers may negatively impact the performance of neighboring ESs. Specifically, such an ES would offload excess requests to nearby servers, thereby consuming their storage capacity and potentially displacing locally important content.

(b) In contrast, a fully private storage scheme would inhibit the ability of ESs to quickly adapt to system-wide content popularity trends. This lack of collaboration limits the caching efficiency, especially in dynamic environments with fluctuating user demands.

To identify a good value of partitioning ratio, we evaluated our approach using real-world datasets that exhibit realistic user request frequency distributions, enabling us to empirically assess and refine the strategy for maximum performance gains.
 
We summarize our key contributions as follows:
\begin{itemize}
    \item We formulate a novel profit maximization problem that jointly integrates the principles of \textit{collaborative caching}, \textit{erasure coding}, and \textit{elastic caching with dynamically adjustable private space} in edge cloud systems. To the best of our knowledge, this is the first work to explore all three concepts in a unified framework.

    \item We are the first to address profit maximization in data access systems utilizing erasure coding, marking a significant step forward in this domain.

    \item We propose an efficient data replacement policy that dynamically adapts to changing access patterns and storage constraints. While earlier works such as \cite{Roy} have considered static partitioning, our approach introduces dynamic adaptation based on the incoming request rate at each edge server—an aspect not addressed previously.

    \item We conduct a comprehensive comparative evaluation of our proposed DSPE strategy against three state-of-the-art baselines: (i) DCC (Distributed Co-operative Caching), (ii) DSP (Dynamic Space Partition), and (iii) E (Erasure-Coded Data Items). Experimental results show that our method consistently outperforms these baselines in terms of overall system profit across diverse workload scenarios.
\end{itemize}

 \section{Related Work} \label{RW}
 Several prior works have explored various aspects of edge caching and profit optimization in edge computing systems, each addressing distinct components of the problem space.

Xiao \textit{et al.} \cite{Xiao21} explored cooperative caching in mobile cloud-edge environments, leading to better service availability, reduced network congestion, and improved user experience, but their model overlooked storage management from a profit maximization standpoint, leaving system efficiency partially unaddressed.
In contrast, Zhao \textit{et al.} \cite{Zhao23} addressed profit maximization in cache-aided intelligent computing networks by proposing an optimization framework that jointly manages caching and computing resources to enhance operator revenue. Their simulations showed gains in resource efficiency, congestion reduction, user satisfaction, and profitability; however, they did not consider collaboration among edge servers.

Somesula \textit{et al.} \cite{manoj} proposed a cooperative service placement and request routing scheme for mobile edge networks to support latency-sensitive applications. Simulations show improved performance over existing methods in terms of delay, cache hit ratio, and cloud load. However, it assumes static user behavior and service popularity, and does not fully address mobility, energy efficiency

Ren \textit{et al.} \cite{a1} introduced a cooperative edge caching framework using “cooperative caching regions” to reduce caching density and promote local data access, lowering dependence on cloud services. While cost-effective, the approach lacks support for erasure coding and storage elasticity.
Xu \textit{et al.} \cite{b8} explored service caching in mobile edge clouds, proposing a distributed mechanism that ensures non-negative gains for all service providers. However, from the edge server perspective, the approach suffers from load imbalance and higher data redundancy.
Luo \textit{et al.} \cite{Luo} proposed a de-duplication strategy to reduce data redundancy by retaining content on edge servers serving the most users within a latency bound. While effective in minimizing replication, the approach overlooks storage costs, limiting its focus on overall resource efficiency and profitability.

\begin{table}[tb!]
\footnotesize
\setlength{\tabcolsep}{0.8mm}
    \centering
    \begin{tabular}{|c|c|c|c|c|c|c|}
    \hline
    Paper & ON/OFF Model & Deadline & C & ER & EL & OBJ \\
    \hline
    Zhao \textit{et al.} \cite{Zhao23}& OFF & \cmark &\xmark & \xmark & \xmark & Profit \\
   Ren \textit{et al.} \cite{a1}& ON & \xmark &\cmark & \xmark & \xmark & Profit \\
   Jin \textit{et al.} \cite{jin2022cost}& ON & \xmark & \xmark & \cmark & \xmark & Storage cost \\
   
    Huang \textit{et al.} \cite{a2}& ON & \xmark & \xmark & \cmark & \xmark & Storage cost \\
    
   Priscilla \textit{et al.} \cite{priscilla}& OFF & \xmark &\cmark & \xmark & \cmark & VM migration cost \\
  Roy \textit{et al.} \cite{Roy}& ON & \cmark &\cmark & \xmark & \cmark & Profit \\
   
   Our work & ON & \cmark & \cmark & \cmark & \cmark & Profit \\
    \hline
    \end{tabular}
    \caption{Summary of existing literature. ``OFF'' and ``ON'' denote offline and online approaches, respectively. ``D'' indicates deadline-aware scheduling, ``C'' represents collaborative caching, ``ER'' refers to erasure coding, ``EL'' stands for elastic caching, and ``OBJ'' indicates the optimization objective considered.}
    \label{tab:lit_rev}
\end{table}

Li \textit{et al.} \cite{Guo} proposed an edge–cloud collaborative offloading strategy for mixed traffic with different delay requirements. It proposes the DGEM algorithm to balance energy consumption and delay guarantees using Lyapunov theory. The approach improves performance over baselines in simulations. However, it relies on simplified assumptions, lacks real-world testing, and doesn't address scalability or learning-based adaptability.

Huang \textit{et al.} \cite{a2} applied erasure coding in the Windows Azure Storage system to minimize the number of fragments needed for data reconstruction while keeping storage overhead low. Their approach achieved a storage overhead of just 1.33×—notably lower than traditional triple replication—while maintaining high data durability.
Jin \textit{et al.} \cite{jin2022cost} proposed EC-EDP, an erasure coding-based edge data placement method to reduce storage costs over replication. While effective in saving space, it assumes static access patterns and homogeneous edge nodes, lacking adaptability, QoS considerations, and fault tolerance under dynamic conditions.

In our prior work \cite{Roy}, we proposed a split-storage model for edge servers with private and shared spaces, along with a Zipf’s Law-based multi-factor replacement policy to maximize profit. Experiments on real-world datasets showed up to 22\% profit gain over state-of-the-art methods. However, the model lacks adjustment of private storage in response to varying AP request loads per ES.

For ease of reference, we summarize the literature review in \autoref{tab:lit_rev}. This work advances existing literature by integrating erasure coding with a dynamic storage partitioning strategy that divides each edge server’s storage into private and public regions. Unlike prior approaches, our design adapts to real-time request patterns, enhancing data availability, reliability, and overall storage efficiency in edge environments.

\section{System Model and Problem Formulation}
\label{sys_mod}

\begin{figure}[tb!]
  \centering
  \includegraphics[width=0.49\textwidth]{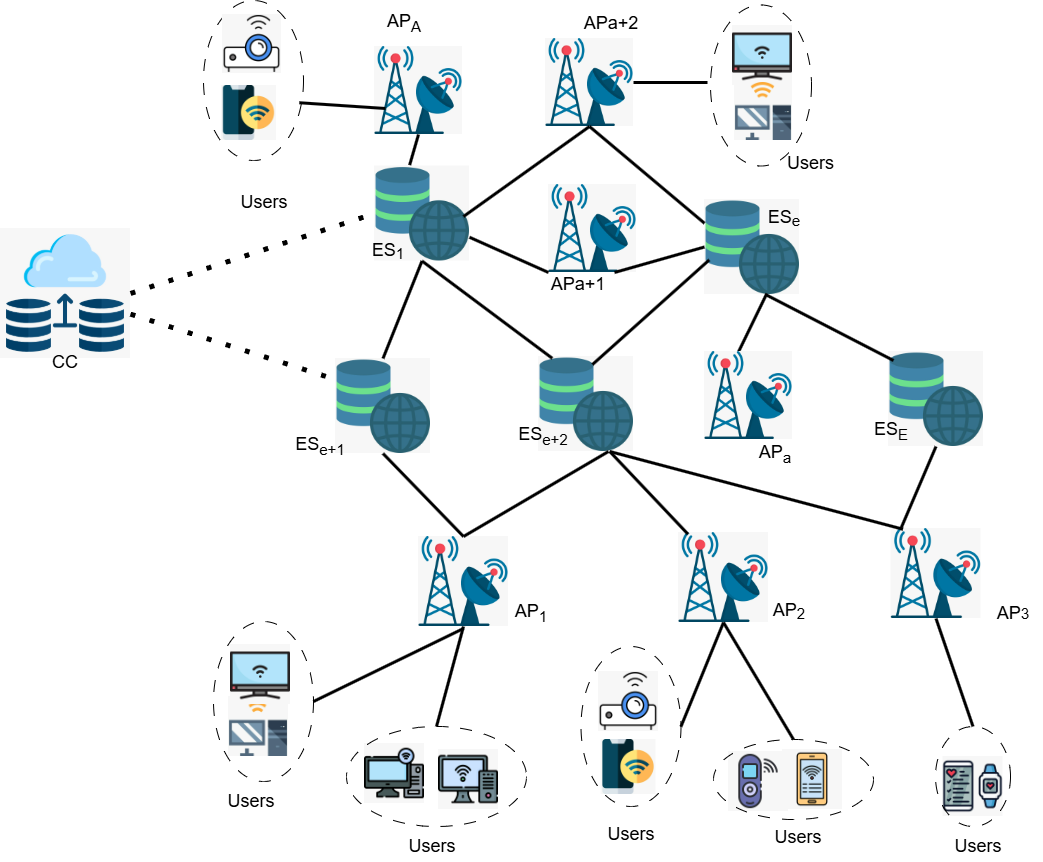}
  \caption{Architecture of Edge-Cloud System Model.}
  \vspace{-0.4cm}
  \label{fig:arc}
\end{figure}

\autoref{fig:arc}, considers a storage cloud network architecture within an edge–cloud system. In our architecture, users access data by submitting requests through Access Points (APs). These APs serve as intermediaries between users and Edge Servers (ESs), which store the requested data. Notably, the APs are lightweight, stateless nodes with no inherent storage capabilities. Their primary role is to facilitate low-latency communication by efficiently collecting and forwarding user requests to the appropriate edge servers.

\begin{table}[tb!]
  \small
  \centering
  \caption{Notation and Description of symbols used}
  \label{tab1}
  \footnotesize
  \begin{tabular}{|m{17mm}|m{63mm}|}
    \hline
    \textbf{Notation} & \hspace{2cm}\textbf{Description}   \\
    \hline
    \centering $CC$ & Centralized cloud   \\
    \hline
    \centering $E$, $A$, $N$ & Number of edge servers (ESs), access points (APs), user requests respectively  \\
    \hline
    \centering $ES_e$, $AP_a$  & $e^{th}$ edge server, $a^{th}$ access point  \\
    \hline
    \centering$\lambda_{AP_a}$  & user request arrival rate at $AP_a$ \\
    \hline
    \centering $S_{e}$, $S_{CC}$,  & Capacity of $ES_e$ and CC respectively \\
    \hline
    \centering $\alpha : \beta$ & Proportion of private storage space : Proportion of shared storage space \\ 
    \hline
     \centering$C_{size}$, $H$ & Size and total number of data and/or parity blocks in the system\\
    \hline
     \centering$R_i$, $X_i$, $d_i$, $\Psi_i$ & $i^th$ Request, content required by, deadline and revenue from $R_i$ respectively\\
    \hline
    \centering $D^{EE}$, $D^{AE}$, $D^{EC}$ & Distance matrix for ES-ES distance, AP-ES distance, and ES-CC distance respectively \\
    \hline
    \centering $P^{AE}[a][e]$ & Probability of $AP_a$ selecting $ES_e$ \\
    \hline
     \centering $K$, $M$ & data and parity blocks respectively \\
    \hline
    \centering $\gamma_a$ &  time required per hop from AP to ES \\
    \centering $\gamma_e$ &  time required per hop from ES to ES \\
    \centering $\gamma_C$ &  time required per hop from CC to ES \\
    \hline
    \centering $T_{ae}$ &  time required to fetch a coded block from  $ES_e$ to $AP_a$ \\
    \hline
    \centering $T_{aec}$ &  time required to fetch a coded block from CC to $AP_a$ via $ES_e$\\
    \hline
  \end{tabular}
\end{table}

\subsection{System Model}\label{SysModel}

As shown in \autoref{fig:arc}, we consider a system comprising $E$ Edge Servers (ESs), each provisioned with an equal amount of data storage capacity. The ESs are indexed with $e$, where $ES_1$ refers to the first ES, and so on upto $E$. The set of ESs is denoted as $  \{ES_{1}, ES_{2}, \dots, ES_{e}, \dots, ES_E\}$. These edge servers are hierarchically distributed across multiple levels within the network and are sparsely interconnected. A subset of these ESs maintains direct connections to a Centralized Cloud (CC), which holds the complete dataset. Due to the limited storage capacity of edge servers, only a fraction of the total data can reside at the edge at any given time.

Each edge server is associated with a set of Access Points (APs), which serve as the interface between end-users and the edge infrastructure. The APs are indexed by $a$ and denoted as $AP_a$, where $1 \leq a \leq A$. These access points are responsible for aggregating user requests and forwarding them to their respective connected ESs. Each AP, in turn, is linked to a group of users, enabling low-latency data access at the network edge.

\subsection{User Requests}

Each user initiates a data access request, referred to as a user request, by submitting it to the Access Point (AP) to which they are directly connected. The data stored in the Centralized Cloud (CC) is partitioned into uniform-sized chunks, denoted by $C^{\text{size}}$. Users request the retrieval of these data chunks from the Edge Servers (ESs). While the CC maintains the complete dataset, selected chunks are strategically cached at the ESs based on a predefined content replacement strategy.

A user request, denoted by $R_i$, is represented as a tuple comprising the access point at which the request is submitted ($AP_a$), the specific content requested ($X_i$), the request deadline ($d_i$), and the revenue associated with the successful completion of the request ($\Psi_i$). Formally, it is defined as:
\begin{center}
$R_i = (\ X_i,\ AP_a,\ d_i,\ \Psi_i)$
\end{center}

Each user is statically assigned to a single, geographically proximate AP, ensuring low-latency communication. Consequently, all user requests are routed through their respective APs to retrieve the requested data items from the underlying edge infrastructure.

\subsection{Distance Model}

The spatial positioning of ESs and APs plays a critical role in determining the overall cost and efficiency of the storage cloud system. Notably, prior studies have shown that the search time for locating erasure-coded blocks is negligible compared to the actual data retrieval time from the ESs\cite{kurata, b5}. In the context of our work, we do not impose any upper bound on the maximum hop distance an edge server is allowed to search within the network. This design choice enables greater flexibility in content retrieval, potentially improving data availability and reducing redundant storage overhead.

To represent the distances between ESs, we define a two-dimensional matrix, $D^{EE}$, where $D^{EE}[e_1][e_2]$ signifies the distance between $ES_{e_1}$ and $ES_{e_2}$. Similarly, we introduce the matrix $D^{EC}$, indicating the distances between the CC and ESs. $D^{EC}[e_1]$ represents the distance between $ES_{e_1}$ and CC if a connection exists.
Additionally, we define another two-dimensional matrix, $D^{AE}$, to store the distances between an AP and ES. $D^{AE}[a][e_1]$ represents the distance between $AP_{a}$ and $ES_{e_1}$. In cases where $AP_{a}$ and $ES_{e}$ are not connected, the distance value is set to $\infty$. Request from an AP can be forwarded only to the ESs the AP is directly connected to. In our model, we assume that the values of the three distance matrices remain constant and are updated only over long time intervals, reflecting relatively stable network conditions. Specifically, the entries in the matrix ($D^{EC}$) are assumed to be an order of magnitude larger than those in matrix ($D^{EE}$). Furthermore, the values in $D^{EE}$ are considered to be higher than those in matrix ($D^{AE}$) for directly connected ES–AP pairs, consistent with empirical observations reported in\cite{xia21}. This hierarchical distance modeling captures the relative latency and communication costs across different layers of the edge–cloud architecture.

\subsection{Storage Model}

Reed-Solomon (RS) code is a class of forward error correction (FEC) code based on polynomial interpolation, widely used for correcting erasures in over-sampled data streams \cite{Rui}. We use erasure code to store data blocks in the cloud network. Erasure code uses RS code as a fundamental building block \cite{Yupeng}. For instance, consider a scenario where a 1 GB data object is partitioned into 10 blocks of 100 MB each. In such a case, the corruption or loss of even a single block can compromise the recoverability of the entire dataset.

\begin{figure}[tb!]
    \centering
    \includegraphics[width=0.4\textwidth]{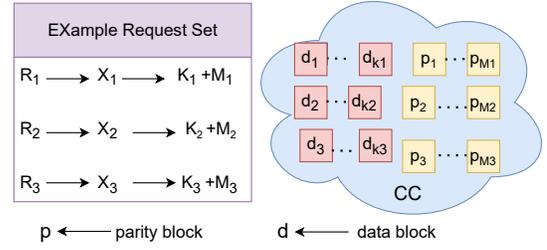}
    \caption{Cloud Storage Model }
    \label{fig:storage1}  
\end{figure}

\begin{figure}[tb!]
    \centering
    \includegraphics[width=0.4\textwidth]{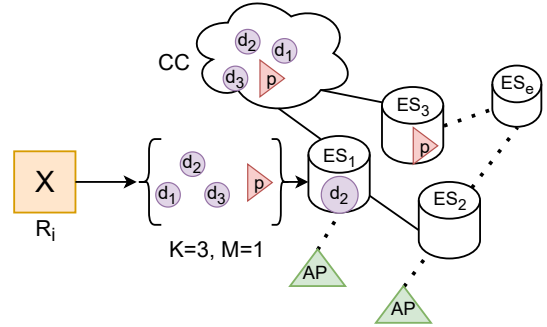}
    \caption{Erasure Split Storage Model }
    \label{fig:storage}  
\end{figure}

Within the erasure coding framework, a content $X$ is segmented into 
$K$ data blocks, and $M$ additional parity blocks are generated to produce a total of $(K+M)$ encoded blocks, as illustrated in \autoref{fig:storage1} and \autoref{fig:storage}. To successfully reconstruct the original data $X$, a user must retrieve any 
$K$ distinct blocks, comprising either data or parity blocks. Although blocks are logically categorized as data or parity, all encoded blocks are of equal size.

In both CC and ESs, data items are stored entirely as encoded blocks. The storage overhead is an important metric used to quantify the additional storage required in comparison to traditional replication. It is formally defined as the ratio between the total number of stored blocks and the number of original data blocks, i.e.,
\begin{equation}
   Overhead = \left(\frac{K + M}{K}\right)
\end{equation}

The encoding and decoding phase is formally defined with the below equations:
\begin{align}
\text{$Q$} &= g \cdot X \hspace{0.2
     cm}and \hspace{0.2 cm} 
     X = g^{-1}\cdot Q
\end{align}   
where, X is the message block and X $\in$ $F^K$, Q is the encoded block and Q $\in$ $F^{(K+M)}$, $g$ is the generator function and $g$ $\in$
$F^{K*(K+M)}$. $F$ is a finite field as mentioned in \cite{khatami}.

Consider a 1 GB movie that must be accessible to users within an Edge Storage System (ESS). To guarantee availability under a traditional replication-based 
scheme, the movie 
would be duplicated across at least two edge servers, resulting in a total storage requirement of 2 GB. However, erasure coding can significantly reduce this overhead. For instance, encoding the movie into 2 data blocks and 1 parity block (each of size 0.5 GB) requires only 1.5 GB of distributed storage. This configuration allows the original movie to be reconstructed from any 2 out of the 3 coded blocks, assuming access within a maximum of two network hops.

Alternatively, a (3,1) erasure coding configuration consisting of 3 data blocks and 1 parity block, each of size approximately 0.33 GB further reduces the total storage to approximately 1.33 GB. In contrast, a (4,4) configuration increases the storage demand to 2 GB. Among these schemes, the (3,1) encoding achieves the lowest storage overhead.
Furthermore, an edge server can process multiple requests in parallel and retrieve the required $K$ coded blocks concurrently if they are available. This parallelism enhances data accessibility and reduces retrieval latency in ESS deployments.

Let the total amount of storage in $ES_{e}$ be $S_{e}$. Let $C_{size}$ and $h$ denote size of each block and total number of coded blocks present in $ES_{e}$ respectively. The total storage consumed by all the blocks in $ES_{e}$ should not exceed its
capacity ($S_{e}$), and it is formally represented as:
\begin{equation}
C_{\text{size}} \cdot h \leq S_{e} \quad \forall{e}    
\end{equation}
Similarly, let $H$ denotes the total number of coded blocks in the system and the total capacity of centralized cloud (CC) be $S_{CC}$, which is formally presented
in Equation (2):
\begin{equation}
C_{\text{size}} \cdot H \leq S_{CC}    
\end{equation}

\begin{figure}[tb!]
    \centering
    \includegraphics[width=0.4\textwidth]{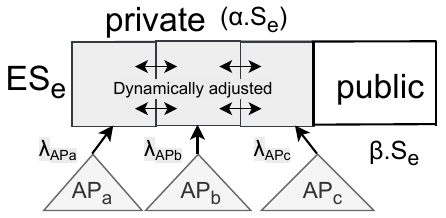}
    \caption{Elastic Storage Model }
    \label{fig:elastic}  
\end{figure}

We divide the storage of an ES into two parts - (a) private
part, (b) shared part. The ratio of the sizes of the two parts is denoted
as $\alpha : \beta$, such that $\alpha + \beta = 1$. The size of the private
part of $ES_{e}$ is $\alpha \cdot S_{e}$, and the size of the shared part of that
ES is $(1 - \alpha) \cdot S_{e}$ as shown in \autoref{fig:elastic}.
We consider a scenario in which an ES is connected to multiple access points APs. In the initial configuration, the ES's private storage space is again partitioned equally among all associated APs. However, if a particular AP exhibits a higher user request rate relative to others, the portion of private storage allocated to that AP dynamically increases to accommodate its elevated demand.

In \autoref{fig:elastic}, $ES_e$ is connected to three APs. Considering the request rate  ($\lambda_a$) at $AP_a$ is higher than the other two, larger portion of the $ES_e$'s private storage is allocated to $AP_a$ in comparison to $AP_{b}$ and $AP_{c}$.

The first part, which is the private (local) space 
of an ES, implies that the blocks there get replaced only to fulfill the content requirements of the connected APs. The second part
of storage is shared (public) space in which blocks get replaced
based on the requirements of the neighboring ESs. 
 An ES is permitted to read data stored in both the public and private storage spaces of its neighboring ESs. However, it can influence it's neighboring ESs to  write only in their public storage space.

\subsection{Real-time Model} 

When a user requests content $X$, via an access point $AP_a$, the data item can be reconstructed by retrieving any 
$K$ out of the total coded blocks (comprising data and parity blocks) stored across ESs. This retrieval mechanism enables the reconstruction of the original content $X$. Upon arrival, each user request can be served through one of two possible approaches:

\textit{Case 1:} Satisfied by ESs only. As shown in \autoref{fig:arc}, an AP is directly linked to certain ESs and indirectly connected to all the ESs. The search proceeds in increasing order of network distance starting from the nearest ES to $AP_a$, and progressing outward. Once an ES, denoted as $ES_{e}$, completes the search for $K$ unique blocks the request is considered satisfied. The response time for this request is defined as the maximum time required to retrieve the 
$K$ blocks from their respective ESs. The actual data retrieval is then initiated, during which the $K$ coded blocks are fetched in parallel from the selected ESs. Time taken to fetch the blocks is calculated based on (hop count $\times$ time$/$hop).

The time taken ($T_{ae}$) is defined by the expression:
\begin{equation}
\text{$T_{ae}$} = (D^{AE}[a][e_1])*\gamma_a + (D^{EE}[e_1][e])* \gamma_e
\end{equation}
\vspace{-0.75 cm}

\begin{equation*}
\text{such that: } \quad 1 \leq a \leq A 
\end{equation*}
\vspace{-0.75 cm}

\begin{equation*}
    1 \leq e_1 \leq e \leq E    
\end{equation*}
\vspace{-0.75 cm}

\begin{equation*}
\gamma_{a} \gets \left( \frac{time}{hop} \right)_{AP\to ES}     \gamma_{e} \gets \left( \frac{time}{hop} \right)_{ES \to ES} 
\end{equation*}

Here $ES_{e_1}$ denotes the initial edge server where the search for the required K coded blocks begins, and 
$ES_e$ represents the final edge server that completes the retrieval process. The search spans a sequence of edge servers, and the path from $ES_{e_1}$ to $ES_e$ traverses the minimum number of intermediate edge servers. This sequence constitutes the shortest path in terms of hop count within the edge network topology.

\textit{Case 2:} If the request is not satisfied by the ESs only, then finally, the remaining coded blocks need to be fetched from the CC.

The time required to traverse a single network hop is influenced by various factors such as network bandwidth, congestion levels, data transfer size, and data type. Here, we model logical distance as a heuristic representation rather than actual physical transmission time, and assume a uniformly provisioned network environment. Therefore, we consider the time per hop to be constant, between (AP–ES), (ES–ES), and (ES–CC) . Here $ES_{e_2}$ denotes the ES through which the remaining blocks are retrieved from the CC into the ES network. 
The time taken ($T_{ae_2c}$) to retrieve coded blocks from CC to $AP_a$ via $ES_{e_2}$ is defined by the expression:

\begin{equation}
\text{Time taken ($T_{ae_2c}$)} =  D^{EC}[e_2]* \gamma_C +T_{ae_2}
\end{equation}
\vspace{-0.25 cm}
\begin{equation*}
\text{$T_{ae_2}$} = (D^{AE}[a][e_1])*\gamma_a + (D^{EE}[e_1][e_2])* \gamma_e
\end{equation*}
\vspace{-0.75 cm}

\begin{equation*}
\text{such that:} \quad 1 \leq a \leq A    
\end{equation*}
\vspace{-0.75 cm}

\begin{equation*}
1 \leq e_1 \leq e_2 \leq E     
\end{equation*}
\vspace{-0.75 cm}

\begin{equation*}
\gamma_{C} \gets \left( \frac{time}{hop} \right)_{CC\to ES}  
\end{equation*}

We have considered, $ \gamma_C > \gamma_e > \gamma_a$.
In all the aforementioned scenarios, profit is accrued only if the total time taken to fulfill the user request does not exceed the specified deadline

\subsection{Problem Formulation}

For each user request that is satisfied within the given deadline, a corresponding revenue is added.

Let \(Z_{i}\) be a binary variable defined as follows:
\[ Z_{i} = \begin{cases} 
1 & \text{if request } R_{i} \text{ is served within the deadline } d_{i} \\
0 & \text{otherwise}
\end{cases}
\]

The revenue associated with request \(R_{i}\) is denoted as \(\Psi_{i}\). The total revenue is the sum of revenue for all requests:

\begin{equation}
    \text{Total Revenue ($\Psi$)} = \max \left( \sum_{i=1}^{\text{N}} Z_{i} \cdot  \Psi_i \right) 
\end{equation}

\begin{thisnote}

So, our goal is to maximize the total revenue of the CDN. We assume a uniform storage cost for each coded block across all ESs and apply a consistent erasure coding scheme for all user requests. The profit is computed as the difference between the revenue generated and the associated storage cost. Given the constant cost assumption, our profit model is primarily formulated in terms of the revenue earned. For clarity and ease of reference, all notations used in this paper are summarized in \autoref{tab1}.

\end{thisnote}

\section{Solution Approach} \label{SolApp}

\begin{thisnote}
In this work, we adopt an online framework for managing data access requests, with the primary objective of maximizing system profit through the integration of erasure coding and elastic caching strategies. The framework is event-driven, triggered upon the arrival of a request at an Access Point (AP).

Once a request arrives at an AP, it initiates the execution of \autoref{algo1}, which invokes the edge server selection mechanism defined in \autoref{alg:edge-server-selection} to determine the most suitable edge server (ES) for offloading the request. The selected ES then handles the request by executing the main request handling framework described in \autoref{algo1_modified}.

Within \autoref{algo1_modified}, the data replacement policy is managed through the algorithm outlined in \autoref{algo2}. Additionally, if a Centralized Cloud (CC) receives the request for data retrieval, the operation is handled using the procedure specified in \autoref{algox}.

\end{thisnote}

Each edge server (ES) maintains a metadata table that plays a vital role in meeting the deadlines of incoming data access requests. This table stores information about the data items currently held at each ES, as well as the network distances between ESs. For an example, the metadata table of $ES_7$ is represented as $(ES_7 <d_{3,8}, d_{2,4}, p_{1,3}>, D^{EE}[e_7][e])$, where $d_{3,8}$ implies that $3^{rd}$ data block of  request $R_8$ is stored in $ES_7$. Similarly, $p_{1,3}$ implies the $1^{st}$ parity block of request $R_3$. $D^{EE}[e_7]$ is the distance matrix for $ES_7$ and all of it's connected ES.
To support efficient retrieval of coded data blocks, the entries in the distance matrix are sorted in ascending order based on proximity to the respective ES.

For simplicity, we assume that metadata updates occur instantaneously, without any delay. Additionally, network parameters such as latency between ESs are considered constant throughout the request handling process.

\subsection{Intuition behind the proposed approach}

\begin{thisnote}

The proposed approach leverages a key property of dynamic space partitioning of the edge servers and erasure code technique for effective reconstruction of the required data blocks.
\begin{enumerate}
    \item Dynamic Space Partitioning:  If the space was fully public, then an ES having a huge influx of requests with respect to its peers would affect the space of other ESs. On the other hand, making the ES fully private the ESs can not take early advantage of global trends of a popular data chunk. 

   Additionally we consider that the private space is also partitioned among the APs connected to the ES and is dynamically adjusted over time. APs with low demand do not unnecessarily hold private storage, which can instead be better used by higher-demand APs. This allows for scalable performance as demand changes over time or due to location-specific spikes and also helps in preventing overload on certain APs.
   
   \item Erasure Coded Data: In our approach we use erasure coding to encode the content block into equal size erasure coded blocks. It ensures data remains recoverable despite failures of some blocks. For example, assume a content of 1 GB size is divided into $10$ equally sized data blocks of 100 MB each. Without replication, failure of a single block makes the entire data unrecoverable. The redundancy ratio in this case is $\frac{10}{10} = 1$. Whereas, using EC(10,4) scheme data can be recovered from any $10$ out of the $14$ coded blocks resulting a higher redundancy ratio of $\frac{14}{10} = 1.4$, making the system more fault tolerant. 
   Using standard replication (e.g., 2x or 3x) storage space of 2 GB and 3GB is required for the two cases respectively. Whereas,  using EC(10,4) scheme only $(10 \times 100 MB + 4\times 100 MB)= 1.4 GB$ of storage space gets utilized.

\end{enumerate}

\end{thisnote}  

\subsection{Request Handling in Access Point}

A user data access request ($R_i$) reaches an access point $AP_a$ to which it is connected to for accessing content ($X_i$). The $AP_a$ checks the metadata table containing information about ESs holding specific coded blocks. The search for coded blocks is explained with the example shown in figure 

In \autoref{fig:example}, there are eight ESs distributed across the network, while being sparsely connected among themselves. There are five access points (AP), and all the ESs are connected to one or more APs. An user request $R_i$ arrives at $AP_1$ with a request for data content $X_i$. We assume that, content $X_i$ is divided into \textit{EC(3,1)} scheme where \textit{K=3 and M=1}. $R_i$ needs to access any of the three coded blocks, in order to reconstruct $X_i$. An AP can choose any of it's connected ES using \textbf{Edge Server Selection},i.e.\autoref{alg:edge-server-selection}. In \autoref{fig:example} we, assume that $AP_1$ chooses $ES_6$ using \autoref{alg:edge-server-selection}.

\subsection{Request Handling in Edge Server}
The search for coded blocks begins with $ES_6$. 
Only two of the three coded blocks are present on $ES_7$ and $ES_2$, while CC contains all of them. The next steps are as follows:
\begin{enumerate}
    \item $ES_6$ refers to distance matrix $D^{EE}$, and starts the search for coded blocks starting with $ES_7$ located at 1 hop distance and $ES_7$ returns only one coded block $d_1$.
    \item $ES_4$ and $ES_8$ are at 2 hop distance and unable to return any coded blocks.
    \item $ES_2$ is at 3 hop distance, and returns $d_2$.
    \item $ES_1$, $ES_3$, and $ES_5$ are at 4 hop distances and unable to return any of the remaining data and/or parity blocks $d_3$ and $p_1$ respectively.
    \item Since, not all K coded blocks are present in the ESs, we adhere to \autoref{algox} for \textbf{Coded Blocks Retrieval from CC}.
    \item Coded block $d_2$ is fetched from $ES_2$ and $d_3$ is fetched from the CC and placed using \autoref{algo2}, i.e. \textbf{Placement of Coded Blocks}.
\end{enumerate}

\begin{figure}[tb!]
    \centering
    \includegraphics[width=0.5\textwidth]{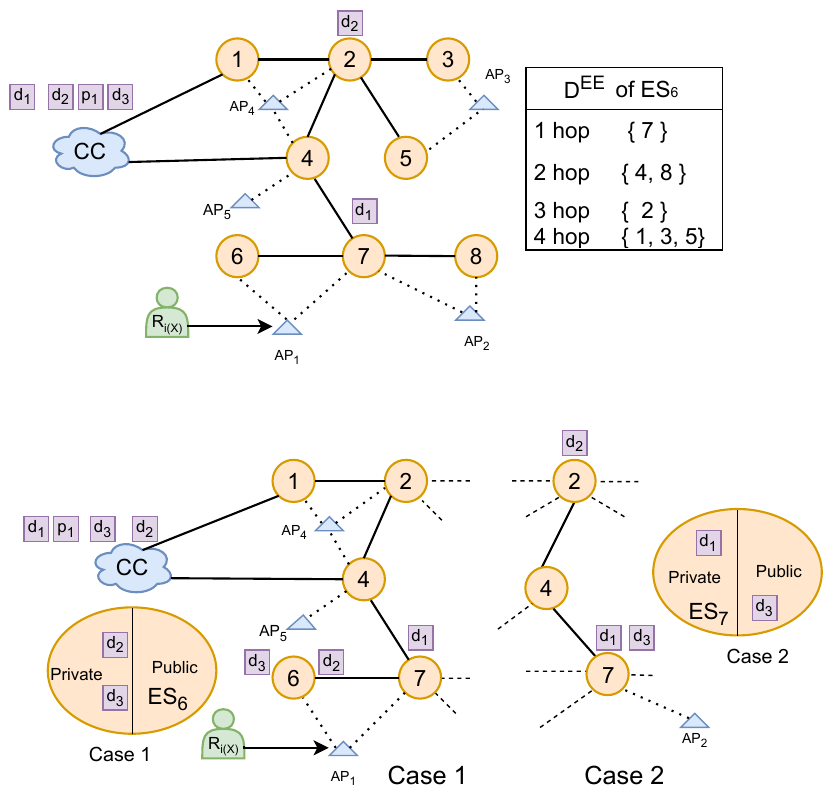}
    \caption{Example demonstrating our solution approach with EC(3,1) and private-public storage partition}
    \label{fig:example}
    
\end{figure}

 Here CC is at maximum distance from $AP_1$ and all request to retrieve coded blocks goes in parallel. So the latency is equal to the time taken to retrieve $d_3$ from CC to $AP_1$. Again, if this latency is less than deadline, then only the corresponding revenue $\Psi_i$ is added to the total revenue. \begin{thisnote}
     
 The coded blocks retrieval process is explained in \autoref{algox}. In \autoref{fig:example}, $ES_4$ is at 2 hop distance from $ES_6$ and also connected to CC. Coded blocks are fetched back into the ES network through $ES_4$.\end{thisnote}

\begin{algorithm}[tb!]
\footnotesize
\textbf{Input:}  $AP_a$ $\gets$  $R_i(X)$

$R_i$ require at least K coded blocks (out of $K+M$) in order to reconstruct X \\
\textbf{Output:} \textcolor{black}{Selects appropriate ES and forwards $R_i$} \\
\textbf{Location:} Access Point ($AP$)
\begin{algorithmic}[1]

\STATE  Request $R_i$ arrives at $AP_a$
\STATE $AP_a$ selects $ES_{e_1}$ using \autoref{alg:edge-server-selection} 
\STATE Forward the request to $ES_{e_1}$ for data access of $X$ (\textit{i.e} it invokes \autoref{algo1_modified})

\end{algorithmic}
\caption{Request Handling in Access Point}
\label{algo1}
\end{algorithm}

\begin{algorithm}[tb!]
\footnotesize
\textbf{Input:}  $ES_{e_1}$ $\gets$  ES set within reachable limits

\textbf{Output:} \textcolor{black}{Schedule $R_i$ to appropriate ESs and CC} \\
\textbf{Location:} Edge Server ($ES$)
\begin{algorithmic}[1]

\STATE $ES_{e_1}$ initiates the search for $K$ coded blocks by identifying the set of edge servers from which data can be fetched for request $R_i$, considering its deadline $d_i$ and the distances defined in $D^{AE}$ (AP-to-ES) and $D^{EE}$ (ES-to-ES)

\IF{any $K$ coded blocks (out of $K+M$) are available within the reachable ES set}
    \STATE Fetch the required coded blocks
    \IF{ $( T_{ae} \leq d_i )$  \hspace{0.2 cm}[ $e \gets$ \textit{farthest ES to return coded blocks} ]}
    \STATE Grant data access to the user through $AP_a$
    \STATE Add revenue $\Psi_i$ from request $R_i$ to the total revenue $\Psi$
    \ELSE { 
    \STATE $R_i$ is rejected and data blocks violating $d_i$ are replaced using \autoref{algo2}}
    \ENDIF

\ELSIF{the missing coded blocks exist outside the reachable ES set}
    \STATE Perform data replacement to bring the required blocks from CC into the reachable set using \autoref{algox}.
    \STATE Coded blocks from CC are replaced using \autoref{algo2}
   \IF{ $( T_{ae_2c} \leq d_i )$  }
    \STATE Grant data access to the user through $AP_a$.
    \STATE Add revenue $\Psi_i$ from request $R_i$ to the total revenue $\Psi$
    \ELSE { 
    \STATE $R_i$ is rejected }
    \ENDIF

\ENDIF

\end{algorithmic}
\caption{Request Handling in Edge Server}
\label{algo1_modified}
\end{algorithm}

\begin{algorithm}[!tb]
\footnotesize
\caption{Edge Server Selection}
\label{alg:edge-server-selection}
\textbf{Input:} $AP_a$, $P^{AE}[a][ES_{AP_a}]$ $\gets$ Probability matrix of $AP_a$ to different ESs connected to $AP_a$\\
\textbf{Output:} Selected edge server ($ES_{e_1}$)\\
\textbf{Location:} Access Point ($AP$)
\begin{algorithmic}[1]
\STATE $AP_a$ refers to the probability matrix $P^{AE}[a][ES_{AP_a}]$ to assess the likelihood of selecting the connected ES
\STATE Based on this probability distribution, $AP_a$ selects one of the ESs connected to AP
\STATE The selected edge server ($ES_{e_1}$) is assigned to handle the request
\STATE \textbf{return} $ES_{e_1}$ to $AP_a$
\end{algorithmic}

\end{algorithm}

\begin{figure*}[tb!]
    \centering
    \includegraphics[width=\textwidth]{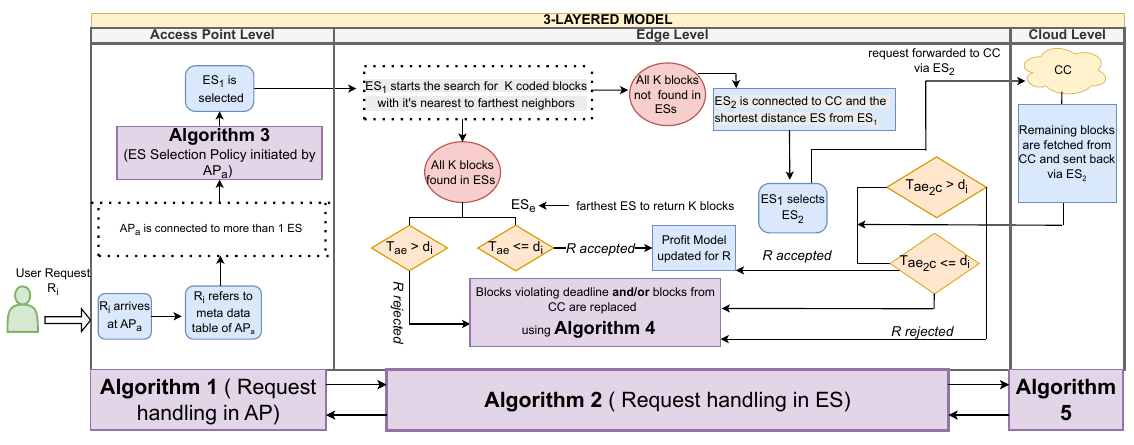}
        \caption{Flowchart of Proposed Approach}
    
    \label{flowchart} 
\end{figure*}

\subsection{Edge Server Selection}
 ES selection policy selects an ES, based on the precomputed access point to edge server probability distribution, thereby guiding the request toward the most probable candidate for successful data retrieval.
To select an ES directly connected to $AP_a$, the system employs a probabilistic selection mechanism based on the precomputed matrix $P^{AE}[a][ES_{AP_a}]$ values.

\subsubsection{AP to ES Probability}

We define \( P^{AE}[A][E] \) as a probability matrix that captures the likelihood of an AP selecting a neighboring ES upon receiving a request. Each entry \( P^{AE}[a][e] \) denotes the probability of \( AP_a \) selecting its neighboring \( ES_e \). The probability values are computed using the following formula:

\begin{equation}
    P^{AE}[a][e] = \frac{1 / D^{AE}[a][e]}{\sum_{ES_e \in \{ES_{AP_a}\}} 1 / D^{AE}[a][e]} 
    \label{eqn:dae}
\end{equation}

Here, \( D^{AE}[a][e] \) represents the distance between \( AP_a \) and \( ES_e \). The term \( \{ES_{AP_a}\} \) denotes the set of all ESs connected to \( AP_a \).
This computation assigns probabilities that are inversely proportional to the distance between \( AP_a \) and each connected \( ES_e \), giving preference to closer servers. The values are then normalized by dividing each inverse distance by the total sum of inverse distances for all ESs connected to the given AP.

For example, if \( AP_a \) is connected to three ESs with distances of 1, 2, and 3 units, respectively, the corresponding probabilities will be: $
P^{AE}[a][e_1] = \frac{1/1}{1/1 + 1/2 + 1/3} \approx 0.54$, $P^{AE}[a][e_2] = \frac{1/2}{1/1 + 1/2 + 1/3} \approx 0.27$, $
P^{AE}[a][e_3] = \frac{1/3}{1/1 + 1/2 + 1/3}\approx0.18$.

This ensures that ESs closer to the AP are selected with higher probability. For ESs that are not directly connected to \( AP_a \), the corresponding probability is set to zero.
The method is to \textit{partition
the interval 
[0,1] into subranges proportional to the values of $P^{AE}[a][ES_{AP_a}]$. A random number $r\in [0,1]$ is generated, and the ES, whose corresponding subrange contains 
$r$, is selected.}

This approach ensures that ESs with higher probabilities are more likely to be selected, optimizing the allocation of resources and improving overall system efficiency.

\begin{algorithm}[tb!]
\footnotesize
\textbf{Input:} \textcolor{black}{coded blocks, $P^{PS} \gets$ hyper-parameter value, edge server ($ES_{e_1}$) }\\
\textbf{Output:} Selected ES for replacement \\
\textbf{Location:} Edge Server ($ES$)
\begin{algorithmic}[1]
\STATE When coded data blocks arrive at $ES_{e_1}$, a decision is made whether to store them locally or forward them
\STATE This decision is influenced by the value of $P^{PS}$, which reflects the probability of retaining blocks privately
\STATE victim selection is done using \textbf{LRU} policy
\IF{the condition favors private storage (based on $P^{PS}$)}
    \STATE The coded blocks are placed in the \textbf{private} region of $ES_{e_1}$
\ELSE
    \STATE A neighboring edge server $ES_{e_1'}$ is selected uniformly at random
    \STATE The coded blocks are stored in the \textbf{public} region of $ES_{e_1'}$
\ENDIF

\end{algorithmic}
\caption{Placement of Coded Blocks}
\label{algo2}
\end{algorithm}

\begin{algorithm}[!tb]
\footnotesize
\caption{ Coded Blocks Retrieval from CC}
\label{algox}
\textbf{Input:} Request for coded blocks not found in ESs   \\
\textbf{Output:} Schedule remaining requests to CC  \\
\textbf{Location:} Centralized Cloud ($CC$)
\begin{algorithmic}[1]

  \STATE $ES_{e_2}$ $\gets$ connected to CC and shortest distance ES from $ES_{e_1}$
  \STATE $ES_{e_2}$ is selected from $D^{EC}[e_2]$ and $D^{EE}[e_1][e_2]$ values
\STATE  Data blocks are fetched from CC via $ES_{e_2}$
\end{algorithmic}

\end{algorithm}

\subsection{Placement of Coded Blocks}
The placement policy is triggered under the following two conditions:

\begin{enumerate}
\item When a user request cannot be satisfied by the available ESs, necessitating the retrieval of the required coded blocks from the CC.
\item When the user request is technically satisfied by the ESs, but the total response time exceeds the specified deadline, thereby violating the system's quality of service constraints.
\end{enumerate}

 \autoref{algo2} shows pseudo code for data placement process. The selection of the ES is determined in a probabilistic manner, considering that neighbouring ESs served requests from nearby areas due to their geographical proximity. This approach proves effective because the content preferences of neighbouring individuals are typically correlated, leading to favourable outcomes. \begin{thisnote}
     
 We use a hyperparameter value $P^{PS}$, which denotes the probability of choosing the ES on which the user request arrives. In our case, we consider $P^{PS} = \alpha$ where $\alpha = $ proportion of private share space in ES. 
 \end{thisnote}The \textbf{victim selection} is done using \textbf{Least Recently Used (LRU)} policy.
 The placement policy can be explained with the example shown in \autoref{fig:example}, assuming that $C_1$ is by default stored in public space of $ES_7$. 
 \begin{enumerate}
     \item \emph{Case 1: Coded block $c_2$ cannot be served within deadline, and the selected ES is $ES_6$}

     Since, $c_3$ is not present in the ESs, and $c_2$ cannot be served within deadline, both the blocks are replaced. The selected ES is $ES_6$. Therefore the blocks are placed in the \textbf{private} space of $ES_6$.
     
     \item \emph{Case 2: Coded block $c_2$ is served within deadline, and the selected ES is $ES_7$, one hop neighbor of $ES_6$ }

      We consider, $c_1$ is present in \textbf{private} space of $ES_7$. Since, $c_2$ gets served within deadline, only $c_3$ gets placed in the \textbf{public} space of $ES_7$.
     
 \end{enumerate}

\section{Experimental Setup and Result Analysis} \label{EXPAN}
To validate the practical effectiveness of our approach, we conducted extensive experiments. We developed edge-cloud storage simulation environment similar to EdgeCloudSim \cite{sim1}, MintEDGE \cite{sim2} using C++, enabling efficient execution of our methodology.

\begin{table}[tb!]
    
    \scriptsize
    \setlength{\tabcolsep}{0.8mm}
    \centering
    \begin{tabular}{|c|c|}
    \hline
        Parameters & Value/ Range \\
    \hline 
         \# of ESs ($E$) & 30 \\
         \# of APs ($A$) & 100 \\
         Trace size & 1,000,000  \\
         Coded blocks present in ES($h$) & 50 \\
         Number of unique data items & 1000\\
         $K$, $M$ & 10, 4 \\
        $\gamma_a$, $\gamma_e$, $\gamma_C$ & 2-3, 10-15, 25-30 unit, similar to \cite{Roy} \\
        $D^{AE}$, $D^{EE}$,  $D^{EC}$ & Generated by method used in \cite{Roy}\\
        $\alpha$ & 0.7\\
        \hline
    \end{tabular}
    \caption{Values of Parameters used in experiments}
    \label{tab:parameters}
\end{table}

\subsection{Datasets}

\subsubsection{Real-life Dataset}

We conducted our experiments using real-world traces from both the Netflix \cite{netflix_data} and Spotify \cite{spotify_data} datasets. These traces were processed to incorporate user-specific attributes such as location, request deadline, and profit associated with each request.

To determine user locations, we relied on actual user distribution statistics. For example, Spotify serves approximately 675 million users globally, with nearly 30\% based in Europe and operating around 12,000 active servers across various locations \cite{spotify_1}. Similarly, Netflix has over 250 million users distributed worldwide.

User request profits and deadlines were computed using realistic user type distributions and corresponding revenue patterns. For Netflix, we considered its three subscription plans with a profit ratio of 8:10:12, while for Spotify, we used a 10:1 ratio reflecting its two-tier plan structure.

Additionally, to manage limited cache capacity, we adopted the Least Recently Used (LRU) replacement policy in our simulations. Other information about experimental is given in \autoref{tab:parameters}.

\subsubsection{Synthetic Dataset}

To evaluate the generalizability of our approach, we also conducted experiments using a synthetic dataset. The request arrival traces and the proportions of different priorities of requests were generated using a random distribution. The parameter values listed in \autoref{tab:parameters} were used to configure and execute our proposed approach on this dataset.

\subsection{State-of-the-Art Approaches}
The following state-of-the-art approaches are considered for comparison with the proposed method:

\begin{itemize}
    \item \textbf{DCC (Distributed Co-operative Caching):} In this approach, the entire storage capacity of each Edge Server (ES) is treated as a public block, with no application of erasure coding. The data replacement policy at each ES is influenced by data requests originating from both its directly connected Access Point (AP) and neighboring ESs, following the methodology outlined in Ren \textit{et al.}\cite{a1}. This configuration corresponds to a splitting factor $\alpha = 0$.

    \item \textbf{DSP (Dynamic Space Partition):} This method partitions the ES storage into private and public regions without employing erasure coding. The division ratio between these regions is inspired from the strategy proposed in Roy et al. \cite{Roy}.

    \item \textbf{E (Erasure-Coded Data Items):} In this scheme, ES storage is treated entirely as public, allowing full collaboration across the network. To enhance fault tolerance and data reliability, erasure coding is employed to encode the data blocks. This approach is inspired by the cost-optimized erasure-coded caching model proposed by Jin et al. \cite{jin2022cost}.
\end{itemize}

\subsection{Profit Analysis Across Different Methods with Varied Task Distribution}

\begin{figure}[tb!]
    \centering
    \includegraphics[width=0.5\textwidth]{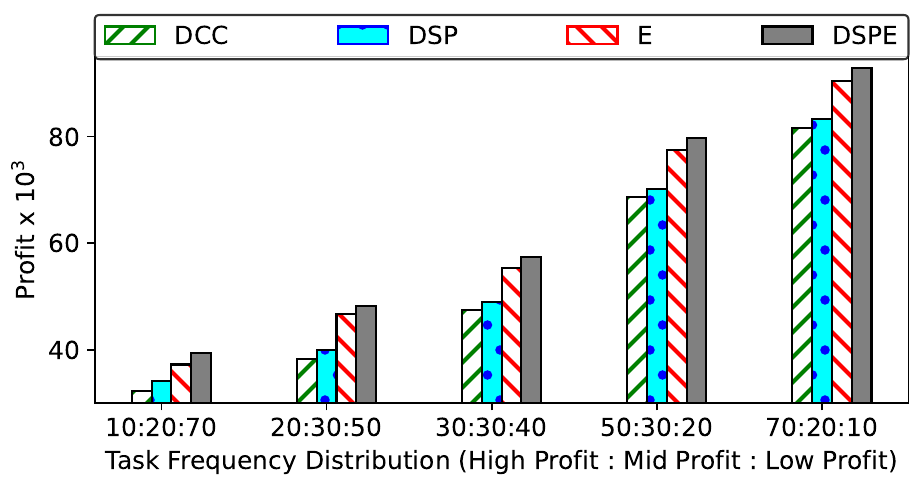}
    \caption{Profit earned using different approaches with varying task distribution ratio for synthetic dataset }
    \label{fig:diff_app_syn}
    
\end{figure}

\begin{figure}[tb!]
    \centering
    \includegraphics[width=0.5\textwidth]{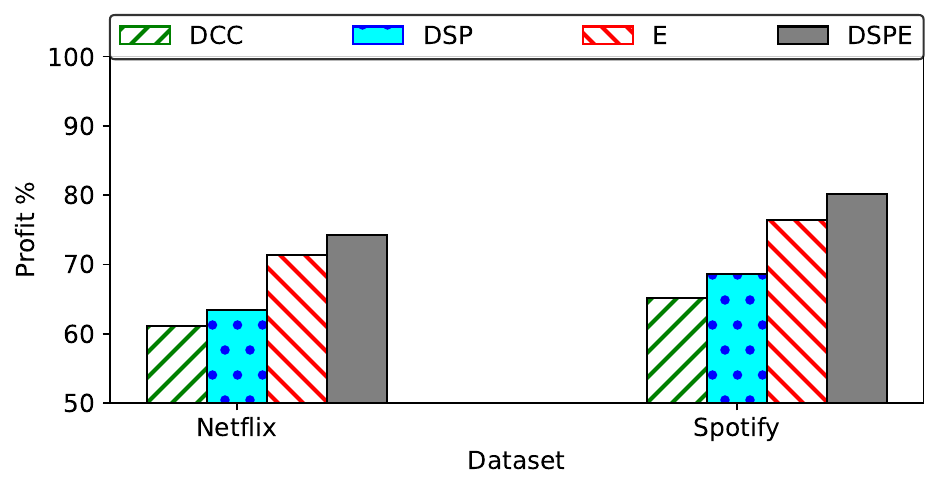}
    \caption{Profit \% of different approaches in real-life datasets}
    \label{fig:diff_app_real}
    
\end{figure}

 \textit{ Observation :} In \autoref{fig:diff_app_syn} we have experimented using synthetic trace with varying task distribution ratio.
 The x:y:z ratio represents the distribution of (high priority : medium priority : low priority) tasks. Various ratios are examined to assess profit outcomes for each method. In \autoref{fig:diff_app_real} we evaluated using both Netflix and Spotify traces, enabling a comparative analysis across different application workloads.
 The experimental results indicate that our proposed approach, \textit{DSPE}, achieves the highest performance among all four schemes.
 In our analysis in \autoref{fig:diff_app_syn}, the 70:20:10 ratio consistently yields optimal results, generating the highest profit.

\textit{Explanation:}  The divergence in profit among methods becomes more pronounced as the distribution skews toward higher priority tasks.

\textit{DSPE} consistently achieves the highest profit, especially in high priority task scenarios. This indicates its superior ability to allocate resources efficiently and prioritize valuable tasks.
\textit{E} and \textit{DSP} show moderate gains, likely due to their design that balances task demands but may not fully exploit the profit potential of high-priority tasks.

\subsection{Effect of Varying Split Ratio of Erasure Code on Profit}

\textit{Observation}: This analysis explores the impact of varying the split ratio of erasure code on profit.

\begin{figure}[tb!]
    \centering
    \includegraphics[width=0.5\textwidth]{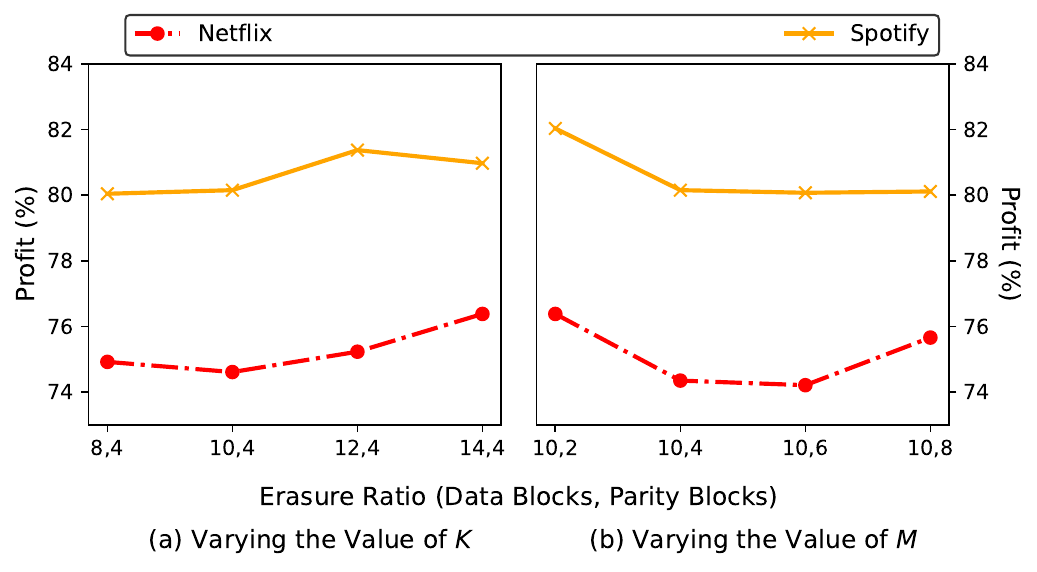}
    \caption{Profit \% of different split ratios of Erasure Code in real-life datasets}
    \label{fig:E_ratio_real}
\end{figure}

 For real life dataset, Netflix and Spotify shown in \autoref{fig:E_ratio_real}, a ratio of 10:2 looks to be the most profitable. For different erasure ratios of $EC(14,4)$ and $EC(10,2)$
 Netflix yields the same profit. Also, erasure schemes containing the same number of coded blocks yields different profits, for example $EC(10,2)$ yields higher profit than $EC(8,4)$ for both Spotify and Netflix.

\textit{Explanation}: 

At $(10,2)$, there is less redundancy, so more storage is available per edge server, allowing more users to be served. But reliability risk is higher, which may cause retransmissions. At $(14,4)$, there is higher redundancy, consuming more space, so fewer users can be served, but with better availability and QoS. These two effects balance out, leading to similar net profit for Netflix.

Secondly, $EC(10,2)$ has a higher erasure ratio $\frac{10}{12} = 0.833$ compared to $EC(8,4) (\frac{10}{12} = 0.667)$. A higher erasure ratio means that a greater proportion of storage is used for original data rather than parity, resulting in more efficient use of cache space. This increase the chances of cache hits and enabling more user requests to be served locally, thereby enhancing overall profit for both Netflix and Spotify.

In conclusion, selecting the optimal split ratio depends on the dataset's characteristics. For large datasets like movies, analyzing data patterns and testing different ratios is crucial to identify the configuration that maximizes profit.

\subsection{Effect of Private Share Proportion on Total Profit}
\begin{figure}[tb!]
    \centering
    \includegraphics[width=0.47\textwidth]{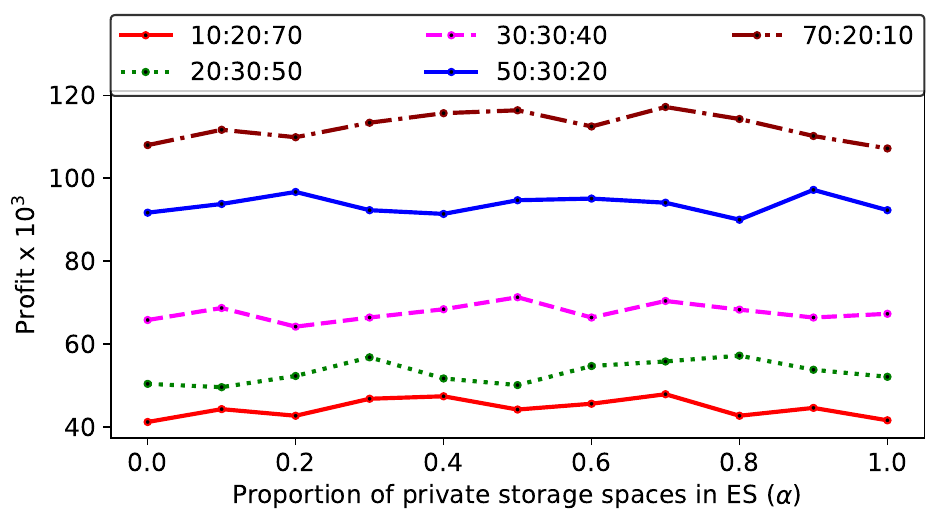}
    \caption{Profit earned by the proposed approach with varying $\alpha$ and task distribution ratio for synthetic dataset }
    \label{fig:alpha_syn}
\end{figure}

\begin{figure}[tb!]
    \centering
    \includegraphics[width=0.5\textwidth]{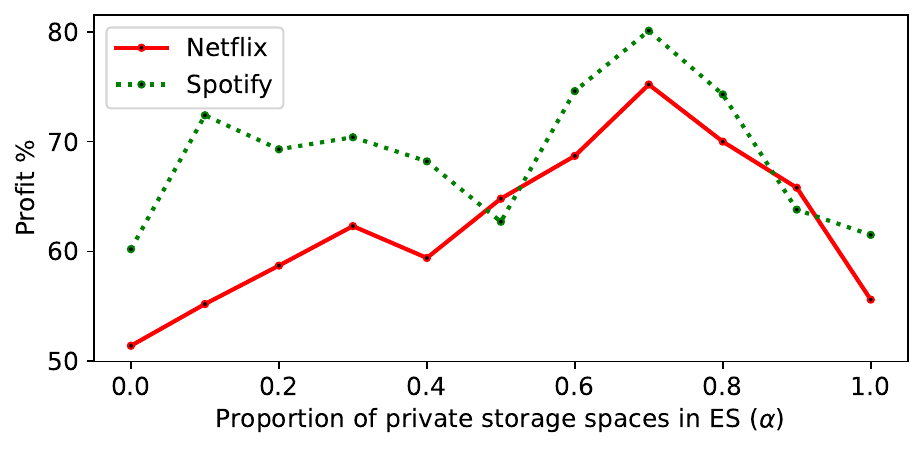}
    \caption{Profit \% of the proposed approach with varying private share
proportion ( the value of $\alpha$) in real-life datasets}
    \label{fig:alpha_real}
\end{figure}

 \textit{Observation:} \autoref{fig:alpha_syn} and \autoref{fig:alpha_real} show the total profit earned as the private share portion increases—\autoref{fig:alpha_syn} focuses on the synthetic dataset with varying task distribution ratios, while \autoref{fig:alpha_real} presents results for both the Spotify and Netflix datasets.

  The value $\alpha$ = 1 indicates that whole ES is private storage space and the value of $\alpha$ = 0 indicates that ES is fully shared (public) \textit{where the ES can influence to replace data
on neighboring ES’s entire storage space.}

This ideal value of $\alpha$ typically ranges from \textcolor{black}{$0.3$ to $0.9$}, depending on the distribution of the trace data. For the synthetic dataset, with the higher number of higher priority requests, the total profit gets maximum at the values of $\alpha$ ranging between $0.7$ to $0.9$.
\textcolor{black}{From \autoref{fig:alpha_syn}, we see that total profit increase as number of high profit tasks increase, this is quite obvious as profit associated with high priority task is higher than the rest. Similarly, for real-life data sets, the total profit is higher when the value of $\alpha$ is $0.7$ for the Spotify data set as well as when the value of $\alpha$ is $0.7$ for the Netflix data set. We choose $\alpha=0.7$ for all our subsequent experiments, as this particular value has the best average result among other values in the mentioned range.}

\textit{Explanation:} 
\begin{enumerate}
    \item \textit{Initial increase in Profit ($\alpha$ from 0 to 0.7) :} As $\alpha$ increases more storage is reserved privately for serving local users directly, reducing dependency on remote or peer ESs. This leads to lower data retrieval latency and reduced transmission costs.
   
    \item \textit{Saturation or decline at High $\alpha$ (1 $\gets$ $\alpha$) :} When almost all storage is private, collaboration among ESs reduces, diminishing the ability to handle content diversity across the network. Less popular content might be evicted due to limited or no shared (public) space. For multi-tenant workloads lack of shared content leads to reduced overall efficiency.
    
\end{enumerate}

\begin{figure}[tb!]
    \centering
    \includegraphics[width=0.48\textwidth]{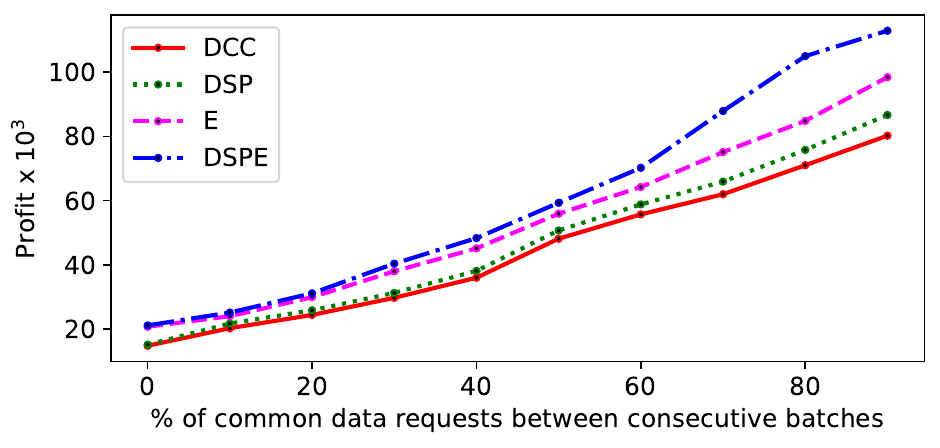}
    \caption{Profit earned by different approaches with varying common data between two consecutive batches}
    \label{fig:data_percent}
\end{figure}

\subsection{Effect of Varying Common Data on Total Profit}
\textit{Observation}: \autoref{fig:data_percent}, shows profit earn when percentage of common data increase. The result is produced with the proportion of high priority/profit, medium priority/profit, and low priority/profit requests as 10\%, 20\% and 70\%. As we can see there is no major difference between $E$ and $DSPE$ when the common data requests is 40\% or less. Similarly, there is also no major difference between $DCC$ and $DSP$ below 40\%. With increase in percentage, $DSPE$ gives the best result as compared to others. i.e. the order of different approaches in terms of profit from highest to lowest is $DSPE$, $E$, $DSP$ and $DCC$ respectively.

\textit{Explanation}:
$DSPE$ outperforms other approaches at high common data ratios by effectively leveraging data reuse and minimizing redundancy. At lower levels of shared data ($\leq$40\%), the benefits of such optimizations are limited since most requests are unique, resulting in similar performance between $E$ and $DSPE$, and between $DSP$ and $DCC$.

\section{Conclusion and Future Work} \label{CNFW}

In this work, we proposed \textit{DSPE}, a profit-driven storage framework for edge-cloud systems that integrates \textit{dynamic space partitioning}, \textit{erasure coding}, and \textit{collaborative caching}. Unlike traditional replication, \textit{DSPE} employs erasure coding for storage efficiency and fault tolerance, while dynamically partitioning each edge server’s storage into private and public regions. The private region is further subdivided among local access points (APs) based on their request arrival rates, enabling responsive, demand-aware storage allocation.

Our framework incorporates a probabilistic server selection mechanism, deadline-aware block retrieval scheduling, and a replacement strategy based on spatial locality and LRU policies. By prioritizing high-value user requests that can be fulfilled within deadline constraints, \textit{DSPE} effectively maximizes cumulative system profit.

For future work, we aim to explore non-uniform storage partitioning across edge servers, support for variable-sized content, and integration of predictive, learning-based prefetching to anticipate content popularity and further reduce latency and cost.

 \setstretch{0.9}
 \footnotesize
 \bibliographystyle{IEEEtran}
 \bibliography{erasure}

 \vskip -2\baselineskip plus -1fil
\begin{IEEEbiography}[{\includegraphics[width=0.7in,height=0.7in]{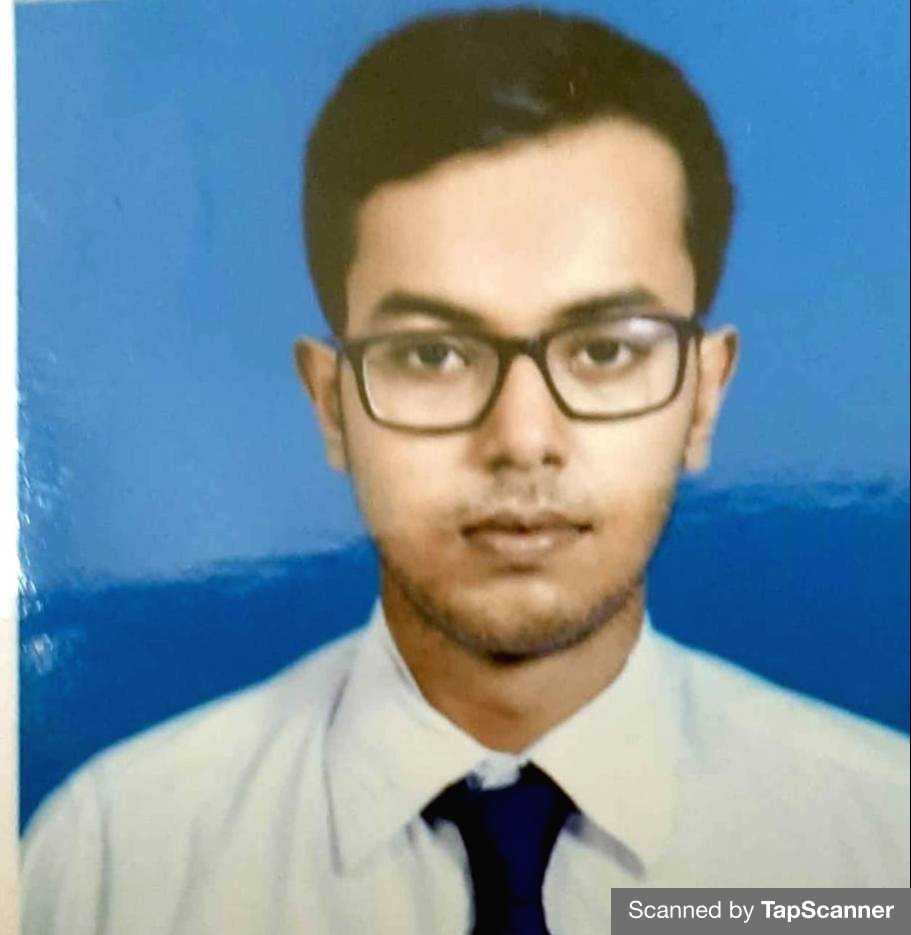}}]{Shubhradeep Roy}
received his B.Tech. degree in Electronics and Communication Engineering from IEM Kolkata. He is currently pursuing his (M.Tech+Ph.D.) Dual Degree in Computer Science at the Indian Institute of Technology Guwahati, India. His research interests include cloud computing, real-time task scheduling, data placement and storage management in edge cloud computing.
\end{IEEEbiography}

\vskip -3\baselineskip plus -1fil

\begin{IEEEbiography}[{\includegraphics[width=0.7in,height=0.7in]{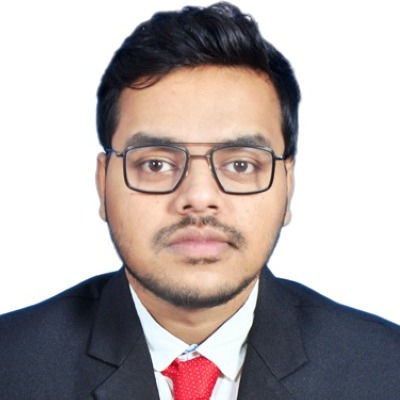}}]{Suvarthi Sarkar}
received his M.Sc. and M.Tech. degrees in Computer Science and Engineering from the University of Calcutta. He is currently pursuing a Ph.D. in Computer Science at the Indian Institute of Technology Guwahati, India. His research interests include cloud computing, real-time task scheduling, containerization, vehicular systems, and vehicular edge cloud computing.
\end{IEEEbiography}

\vskip -3\baselineskip plus -1fil

\begin{IEEEbiography}[{\includegraphics[width=0.7in,height=0.7in]{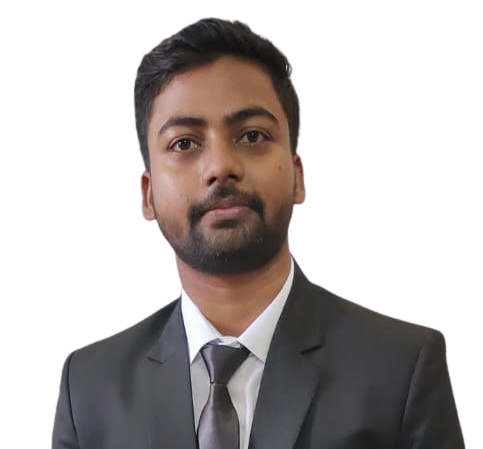}}]{Vivek Verma}
received his B.Tech. in Computer Science and Engineering from Kamla Nehru Institute of Technology, Sultanpur, and his M.Tech. from the Indian Institute of Technology Guwahati in the same field. His research interests include edge-cloud computing, data placement optimization, and erasure coding techniques for large-scale storage systems. He is passionate about developing efficient algorithms for real-world distributed systems.

\end{IEEEbiography}

\vskip -3\baselineskip plus -1fil

\begin{IEEEbiography}[{\includegraphics[width=0.7in,height=0.7in]{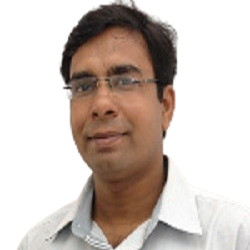}}]{Aryabartta Sahu}
received his Ph.D. in CSE from IIT Delhi. Since 2009, he has been serving as a faculty member in the Department of CSE at IIT Guwahati, India. He has authored over 35 research papers in reputed journals and conferences, including IEEE Transactions, Systems Journal, DATE, CCGrid, HPCC, etc. His research interests span advanced computer architecture, cloud systems, multicore parallel programming and compilation, embedded systems, as well as VLSI and FPGA design. He is actively involved in academic and professional communities and holds membership in ACM and senior membership in IEEE.
\end{IEEEbiography}
\end{document}